\documentclass[conference,twocolumn]{IEEEtran}
\ifCLASSINFOpdf
\else
\fi
\usepackage{amsfonts,amsmath,amssymb}
\usepackage{mathtools}
\usepackage{mdwmath}
\usepackage{cuted}
\usepackage{mdwtab}
\usepackage{amsthm}           
\usepackage{bm}               
\usepackage{units}            
\usepackage{graphicx}
\usepackage{epstopdf}
\usepackage[linesnumbered,ruled,vlined]{algorithm2e}

\usepackage{url}

\usepackage{paralist}
\usepackage{color}
\usepackage{authblk}
\usepackage{cleveref}
\usepackage{float}
\usepackage[caption = false]{subfig}
\usepackage[nolist,nohyperlinks]{acronym} 

\newcommand{\Rs}{\mathbb{R}^2}

\newcommand{\sinr}{\mathrm{SINR}}

\newcommand{\Ru}{v(\gamma,r_s)}
\newcommand{\Rl}{u(\gamma,r_s)}


\begin{acronym}[NC-OFDM]
\acro{3gpp}[3GPP]{3\textsuperscript{rd} Generation Partnership Program}
\acro{5g}[5G]{Fifth Generation}
\acro{Adam}{Adaptive Moment Optimisation}
\acro{ap}[AP]{Access Point}
\acro{atc}[ATC]{Air Traffic Control}
\acro{bs}[BS]{Base Station}
\acro{bpp}[BPP]{Binomial Point Process}
\acro{cc}[C\&C]{Command \& Control}
\acro{cdf}[CDF]{cumulative distribution function}
\acro{cdma}[CDMA]{Code Division Multiple Access}
\acro{cdr}[CD\&R]{Conflict Detection \& Resolution}
\acro{cfo}[CFO]{Carrier Frequency Offset}
\acro{comreg}[ComReg]{Commission for Communications Regulation}
\acro{csi}[CSI]{Channel State Information}
\acro{easa}[EASA]{European Aviation Safety Agency}
\acro{faa}[FAA]{Federal Aviation Administration}
\acro{fso}[FSO]{Free-Space Optical}
\acro{geo}[GEO]{Geosynchronous Equatorial Orbit}
\acro{gps}[GPS]{Global Positioning System}
\acro{gs}[GS]{Ground Station}
\acro{hap}[HAP]{High Altitude Platform}
\acro{iot}[IoT]{Internet of Things}
\acro{kpi}[KPI]{Key Performance Indicator}
\acro{lap}[LAP]{Low Altitude Platform}
\acro{leo}[LEO]{Low Earth Orbit}
\acro{los}[LoS]{Line-of-Sight}
\acro{lte}[LTE]{Long Term Evolution}
\acro{mac}[MAC]{Media Access Control}
\acro{mcp}[MCP]{Matern Cluster Process}
\acro{mc}[MC]{Monte Carlo}
\acro{mimo}[MIMO]{Multiple Input Multiple Output}
\acro{mip}[MIP]{Mixed-Integer Programming}
\acro{mm}[MM]{Mapping Mechanism}
\acro{ml}[ML]{machine learning}
\acro{mno}[MNO]{Mobile Network Operator}
\acro{nlos}[NLoS]{non-Line-of-Sight}
\acro{nn}[NN]{Neural Network}
\acro{ofdma}[OFDMA]{Orthogonal Frequency Division Multiple Access}
\acro{oot}[OOT]{Out-of-Tree}
\acro{osi}[OSI]{Open Systems Interconnection}
\acro{otdoa}[OTDoA]{Observed Time Difference of Arrival}
\acro{ott}[OTT]{Over-The-Top}
\acro{pv}[PV]{photo-voltaic}
\acro{pdf}[pdf]{probability density function}
\acro{ppp}[PPP]{Poisson Point Process}
\acro{qos}[QoS]{Quality of Service}
\acro{rc}[RC]{Remote Control}
\acro{rl}[RL]{Reinforcement Learning}
\acro{rss}[RSS]{Received Signal Strength}
\acro{se}[SE]{Spectral Efficiency}
\acro{sir}[SIR]{Signal-to-Interference Ratio}
\acro{sinr}[SINR]{Signal-to-Interference-and-Noise Ratio}
\acro{snr}[SNR]{Signal-to-Noise Ratio}
\acro{uav}[UAV]{Unmanned Aerial Vehicle}
\acro{ue}[UE]{User Equipment}
\acro{ula}[ULA]{Uniform Linear Array}
\acro{urllc}[URLLC]{Ultra-Reliable Low Latency Communication}
\acro{wsn}[WSN]{Wireless Sensor Network}
\acro{wsn}[WSN]{Wireless Sensor Network}
\acro{rv}[RV]{random variable}
\acro{ppp}[PPP]{Poisson point process}
\acro{pgfl}[PGFL]{point generation functional}
\acro{pdf}[PDF]{probability density function}
\end{acronym}

\begin{document}
%
\title{Intelligent Base Station Association for UAV Cellular Users: A  Supervised Learning Approach}
%
%
%

\author{Boris Galkin*,
        Ramy Amer*,
        Erika Fonseca*,
        and~Luiz~A. DaSilva*$^\dagger$
}

\affil{* CONNECT- Trinity College Dublin, Ireland \\
       $\dagger$ Commonwealth Cyber Initiative, Virginia Tech, USA \\
\textit{E-mail: \{galkinb,ramyr,fonsecae\}@tcd.ie, ldasilva@vt.edu}}

\maketitle

\begin{abstract}
\ac{5g} cellular networks are expected to provide cellular connectivity for vehicular users, including \acp{uav}. When flying in the air, these users experience strong, unobstructed channel conditions to a large number of \acp{bs} on the ground. This creates very strong interference conditions for the \ac{uav} users, while at the same time offering them a large number of \acp{bs} to potentially associate with for cellular service.
Therefore, to maximise the performance of the \ac{uav}-\ac{bs} wireless link, the \ac{uav} user needs to be able to choose which \acp{bs} to connect to, based on the observed environmental conditions. This paper proposes a supervised learning-based association scheme, using which a \ac{uav} can intelligently associate with the most appropriate \ac{bs}. We train a \ac{nn} to identify the most suitable \ac{bs} from several candidate \acp{bs}, based on the received signal powers from the \acp{bs}, known distances to the \acp{bs}, as well as the known locations of potential interferers. We then compare the performance of the \ac{nn}-based association scheme against strongest-signal and closest-neighbour association schemes, and demonstrate that the \ac{nn} scheme significantly outperforms the simple heuristic schemes.
\end{abstract}


\section{Introduction}

\ac{5g} cellular networks are intended to meet the strict latency and reliability requirements of \ac{urllc}, thereby enabling a wide range of new technologies and use-cases. Autonomous vehicles such as self-driving cars and \acp{uav} are predicted to be some of the core users of \ac{5g} networks \cite{Ge_2019}. \acp{uav} are becoming increasingly used in a wide range of applications such as aerial surveillance, safety, as well as product delivery \cite{8660516}. In all of these use-cases, the \acp{uav} require ubiquitous and reliable data communication with their human operators, local authorities, as well as each other, which makes them reliant on the \ac{5g} network.

These \ac{uav} users represent a paradigm shift for the cellular network, as they are able to freely move in three-dimensional space, unlike typical terrestrial users. As they move in the air, \acp{uav} are exposed to vastly different radio environment conditions as compared to terrestrial users, due to the presence of dominant \ac{los} links as well as reduced antenna gains from BS down-tilted antennas \cite{Lin_2017}. 
For instance, \acp{uav} can establish unobstructed \ac{los} wireless links to distant transmitters, which allows them to receive a strong, unattenuated wireless signal from their serving \ac{bs}, but which also make them susceptible to strong \ac{los} interference. 


Recent \ac{3gpp} studies on the performance of cellular connectivity for \ac{uav} users confirm that strong interference in the sky will have a severe impact on UAV users \cite{3GPP_2018}. The studies suggest that this interference can be mitigated by equipping the \acp{uav} with steerable, directional antennas. By steering such an antenna towards the desired serving \ac{bs}, the \ac{uav} can use the strong directional antenna gain to boost the desired signal, while simultaneously attenuating undesirable interfering signals. In our prior work \cite{8422376}, we mathematically model the achievable performance of a \ac{uav} equipped with a steerable antenna which connects to terrestrial infrastructure. Our results demonstrate that \acp{uav} operating at large heights above ground experience strong interference from a large number of \acp{bs} with \ac{los} channel conditions, which can be compensated for via steerable directional antennas, corroborating the conclusions in \cite{3GPP_2018}. In \cite{Amer_2020}, we investigate the handover of \acp{uav} under practical antenna configurations. The authors in \cite{Azari_2018} examine how factors such as \ac{bs} density and height above ground affect the ability of \acp{uav} and ground users to share the network. The same authors in \cite{8692749} extend this by studying the impact of directional antenna tilt and beamwidth. In \cite{Geraci_2018} the authors model the performance of a \ac{uav} with an omni-directional antenna connecting to a \ac{bs} network with \ac{mimo} antenna arrays, and demonstrate how multi-user \ac{mimo} can significantly improve \ac{uav} service quality.

The current state-of-the-art on \ac{uav} cellular communications tends to assume that the \ac{uav} has full awareness of the environment and is therefore able to associate with the most suitable \ac{bs}. Given its aerial position, a \ac{uav} may have a large number of candidate \acp{bs} that it can connect to for cellular service, using its directional antenna. To choose the most suitable \ac{bs} the \ac{uav} needs to be aware of the channel conditions for each candidate \ac{bs}, which involves steering the directional antenna towards each \ac{bs} and assessing the resulting channel quality. Depending on the \ac{uav} use case and the state of the environment, this process may introduce a large overhead to maintaining cellular connectivity, or (in the case of highly dynamic channels when the \ac{uav} is moving) it may not be feasible at all. As an alternative to this iterative channel measurement step, we propose choosing the most suitable candidate \ac{bs} using available environmental knowledge and a trained \ac{nn}.




\acp{nn} have started to gain popularity in the wireless community as function approximators \cite{8755300}. In this regard, the authors in \cite{Chen_2019} have explored the use of supervised learning for training millimeter-wave \ac{mimo} antennas. The authors demonstrate how the observed channel conditions at one antenna can be used to configure an antenna at another location, using a trained \ac{nn}. In \cite{Arvinte_2019}, the authors use \ac{bs} geolocation information to design an \ac{nn}-based scheduler that maximises the system throughput in a millimeter-wave multi-\ac{bs}, multi-user communication scenario. The work in \cite{Alkhateeb_2018} proposes \ac{nn}-based coordinated beamforming, where multiple \acp{bs} simultaneously serve a single user. 

 Our contribution in this paper is to propose an \ac{nn} approach which allows a \ac{uav} user to intelligently select one of the \acp{bs} in the network to associate with, so that the channel quality of the \ac{uav}-\ac{bs} link is maximised. We consider a \ac{uav} equipped with two sets of RF-chains with separate antennas: an omni-directional antenna for measuring the received signal power from nearby \acp{bs}, as well as a directional antenna which the \ac{uav} aligns towards its associated \ac{bs} and uses for data transmission. We train an \ac{nn} to infer which \ac{bs} will give the best channel quality for the directional antenna connection based on the received signal power at the omni-directional antenna, as well as other environmental information that is known to the \ac{uav}. The \ac{uav} is capable of using heuristic \ac{bs} selection strategies; to demonstrate the advantage of our \ac{nn} approach we compare the results against these heuristic strategies, based on both simulations as well as mathematical derivations from our prior works \cite{8422376} and \cite{Amer_2020}. 





\section{System Model} \label{system}

\begin{figure}[t!]
\centering
	\subfloat{\includegraphics[width=.45\textwidth]{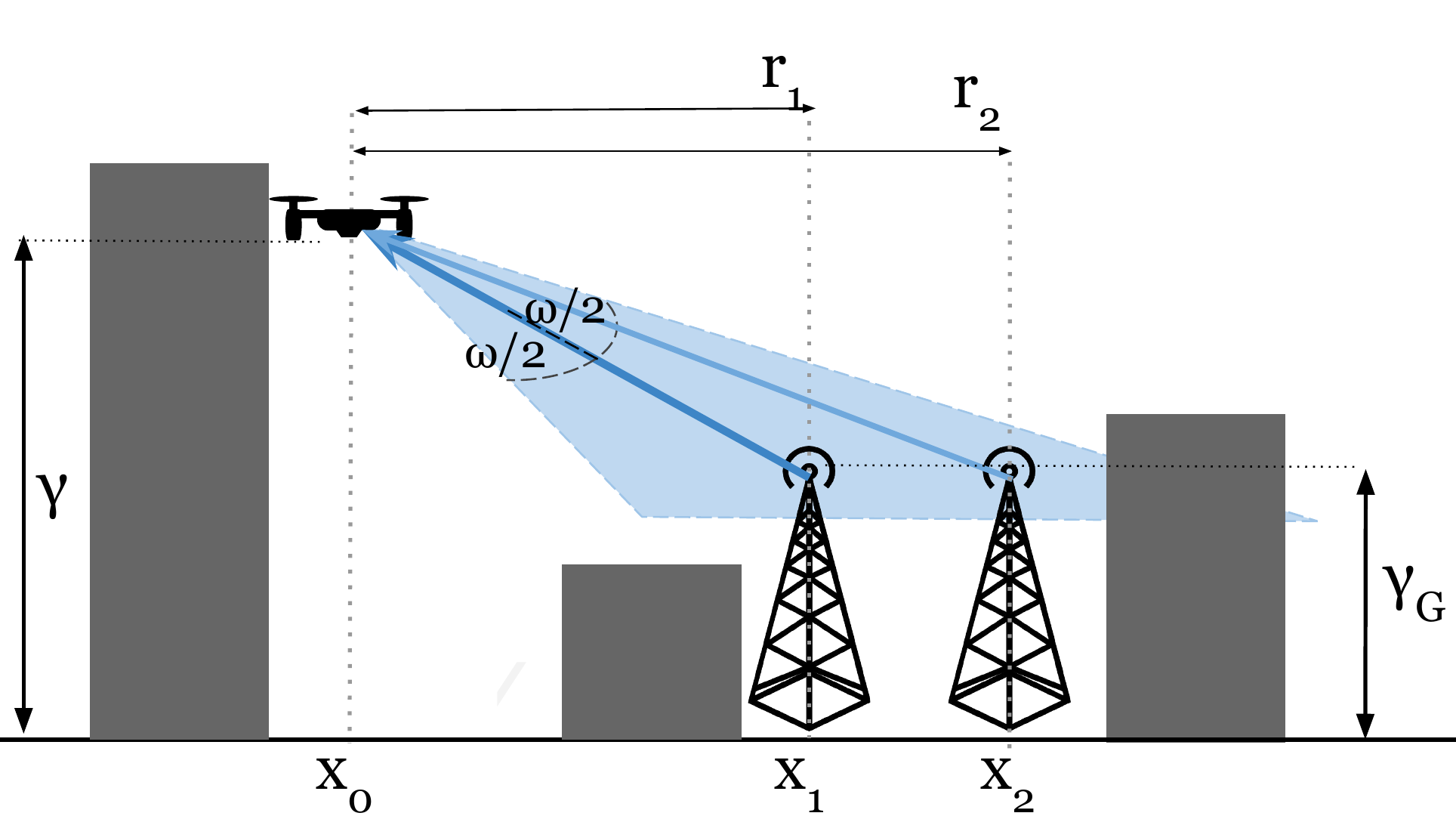}}\\
	\subfloat{\includegraphics[width=.45\textwidth]{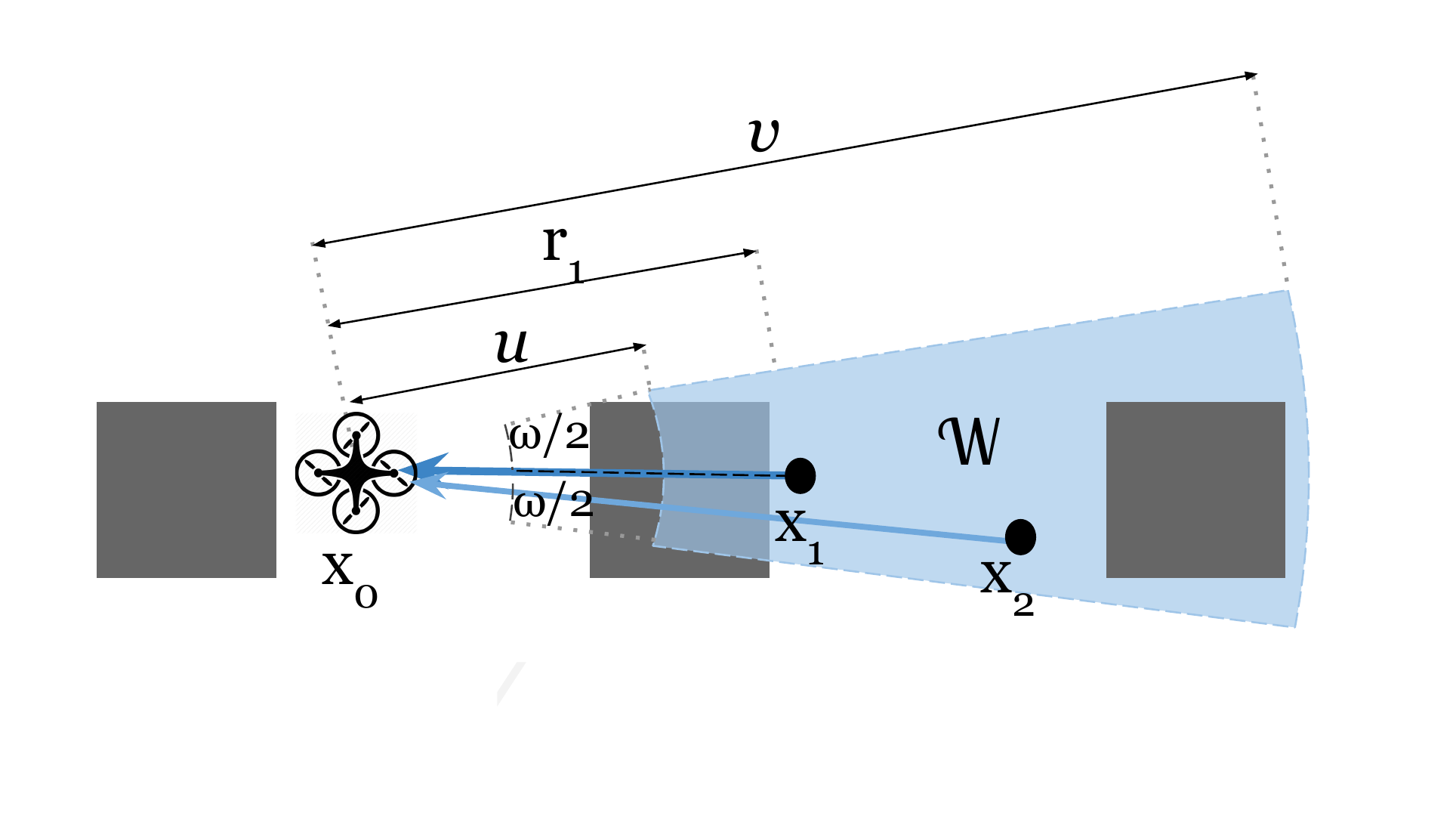}}
	\vspace{-5mm}
	\caption{
	Side and top view showing a UAV in an urban environment at a height $\gamma$, positioned above $x_0$ with antenna beamwidth $\omega$. The UAV associates with the \ac{bs} at $x_1$ and centers its antenna main lobe on the \ac{bs} location; the blue area $\mathcal{W}$ illuminated by the main lobe denotes the region where interferers may be found. The \ac{bs} at $x_2$ falls inside this area and produces interference. 
	\vspace{-5mm}
	}
	\label{fig:drone_network}
\end{figure}


We consider an urban environment where a flying \ac{uav} uses an underlying cellular network for its wireless connectivity, as depicted in \cref{fig:drone_network}. The underlying cellular network consists of \acp{bs} which are horizontally distributed as a homogeneous \ac{ppp} $\Phi = \{x_1 , x_2 , ...\} \subset \Rs$ of intensity $\lambda$, at a height $\gamma_{G}$ above ground. Elements $x_i\in \Rs$ represent the projections of the \ac{bs} locations onto the $\Rs$ plane. The coordinates of the \ac{uav} are denoted as $x_0 \in \Rs$, with the \ac{uav} height above ground denoted as $\gamma$. Let $r_i = ||x_i-x_0||$ denote the horizontal distance between the coordinates $x_0$ and $x_i$, and let $\phi_i = \arctan(\Delta \gamma/r_i)$ denote the vertical angle, where $\Delta \gamma = \gamma - \gamma_{G}$. Without loss of generality we set the horizontal coordinates of the \ac{uav} $x_0$ as the origin (0,0).

The \ac{uav} is equipped with two sets of antennas: an omni-directional antenna for \ac{bs} pilot signal detection and signal strength measurement, as well as a directional antenna for communicating with the \ac{uav}'s associated \ac{bs}. The omni-directional antenna has an omni-directional radiation pattern with an antenna gain of 1, while the directional antenna has a horizontal and vertical beamwidth $\omega$ and a rectangular radiation pattern; following \cite{8422376}, the antenna gain is given as $\eta(\omega) = 16\pi/(\omega^2)$ inside the main lobe and $\eta(\omega)=0$ outside. We denote the coordinates of the \ac{bs} which the \ac{uav} is associated with as  $x_s \in \Phi$ and its horizontal distance to the \ac{uav} as $r_s$. The \ac{uav} aligns its directional antenna towards $x_s$; this results in the formation of an antenna radiation pattern around $x_s$ which we denote as $\mathcal{W} \subset \Rs$, as depicted in \cref{fig:drone_network}. This area takes the shape of a ring sector of arc angle equal to $\omega$ and major and minor radii $\Ru$ and $\Rl$, respectively, where

\begin{align}
\Ru = 
\begin{cases}
\frac{|\Delta \gamma|}{\tan(|\phi_{s}|-\omega/2)} \hspace{-2mm} &\text{if} \hspace{3mm} \omega/2 < |\phi_{s}| < \pi/2 - \omega/2 \\
\frac{|\Delta \gamma|}{\tan(\pi/2 -\omega)} \hspace{-2mm} &\text{if} \hspace{3mm} |\phi_{s}|  > \pi/2 - \omega/2 \\
\infty &\text{otherwise} \nonumber
\end{cases}
\end{align}

\begin{align}
\Rl = 
\begin{cases}
\frac{|\Delta \gamma|}{\tan(|\phi_{s}| +\omega/2)} \hspace{2mm} &\text{if} \hspace{3mm} |\phi_{s}|  < \pi/2 - \omega/2  \\ 
0 &\text{otherwise}
\end{cases}
\end{align}

\noindent
with $|.|$ denoting absolute value. The \acp{bs} which fall inside the area $\mathcal{W}$ are denoted by the set $\Phi_{\mathcal{W}} = \{x \in \Phi : x \in \mathcal{W}\}$. The \acp{bs} in the $\Phi_{\mathcal{W}}$ are capable of causing interference to the \ac{uav}-\ac{bs} communication link, as their signals may be received by the \ac{uav}'s directional antenna with non-zero gain.

In our scenario we consider an urban environment, with a grid of buildings distributed according to a square grid, following the model proposed in \cite{ITUR_2012}. All of the buildings are modelled as having the same square horizontal area, with each building having a Rayleigh-distributed random height. These buildings are capable of blocking the wireless link between a \ac{bs} and the \ac{uav}. To determine if the channel between the \ac{uav} and a \ac{bs} is \ac{los} or \ac{nlos} we carry out a ray-trace; if a building exists between the \ac{uav} and \ac{bs} that is tall enough to block the straight line between the two devices then the channel is considered \ac{nlos}.

\noindent

We assume that the \acp{bs} are equipped with \ac{ula} antennas, with $N_t$ antenna elements. The vertical gain of these antennas is a function of the angle between the \ac{uav} and the \ac{bs} and is defined similar to \cite{8756719} as

\begin{equation}
\label{antten-gain}
\mu(\phi_i) = \frac{1}{N_t}\frac{\sin^2 \frac{N_t \pi}{2}\sin(\phi_i)}{\sin^2 \frac{\pi}{2}\sin(\phi_i)} .
\end{equation}

For simplicity we consider the \ac{bs} horizontal gain to be omni-directional with a value of 1.



\noindent

When the \ac{uav} is connected to the \ac{bs} at $x_s$ and aligns its directional antenna towards it, the \ac{sinr} of the downlink signal received by the directional antenna is given as

\begin{equation}
\sinr = \frac{p H_{t_{s}} \eta(\omega)\mu(\phi_s) c (r_s^2+\Delta \gamma^2)^{-\alpha_{t_s}/2}}{I_{L} + I_{N}+\sigma^2}
\label{eq:SINR}
\end{equation}

\noindent
 where $p$ is the \ac{bs} transmit power, $H_{t_{s}}$ is the random Nakagami-m multipath fading component, $\alpha_{t_{s}}$ is the pathloss exponent, $t_s \in \{\text{L},\text{N}\}$ is an indicator variable which denotes whether the \ac{uav} has \ac{los} or \ac{nlos} to its serving \ac{bs} $x_s$, $c$ is the near-field pathloss, $\sigma^2$ is the noise power, and $I_{L}$ and $I_{N}$ are the aggregate interference from the \acp{bs} which have \ac{los} and \ac{nlos} channels to the \ac{uav}, respectively.
 
 We define an SINR threshold $\theta$ for the wireless link: if $\sinr>\theta$ this represents the \ac{uav} establishing a wireless link to the \ac{bs} at $x_s$ of an acceptable channel quality. We refer to this as the \ac{uav} having coverage from the cellular network.  
 
 
 We assume that the \ac{uav} has the 3D coordinates of the \ac{bs} network $\Phi$, either from a map supplied by the network operator, or through sensing by the \ac{uav} itself. Using this information, in addition to measurements received by the \ac{uav}'s omni-directional antenna, the \ac{uav} makes a decision about which \ac{bs} in $\Phi$ it should associate with, for the purpose of maximising the \ac{sinr} of its communication link. In the next section we describe our proposed supervised learning-based \ac{nn} architecture to carry out this process.
 
 
\section{Machine Learning Approach}

 \subsection{Neural Network Architecture and Configuration}

\begin{figure}[b!]
\centering
	\includegraphics[width=.45\textwidth]{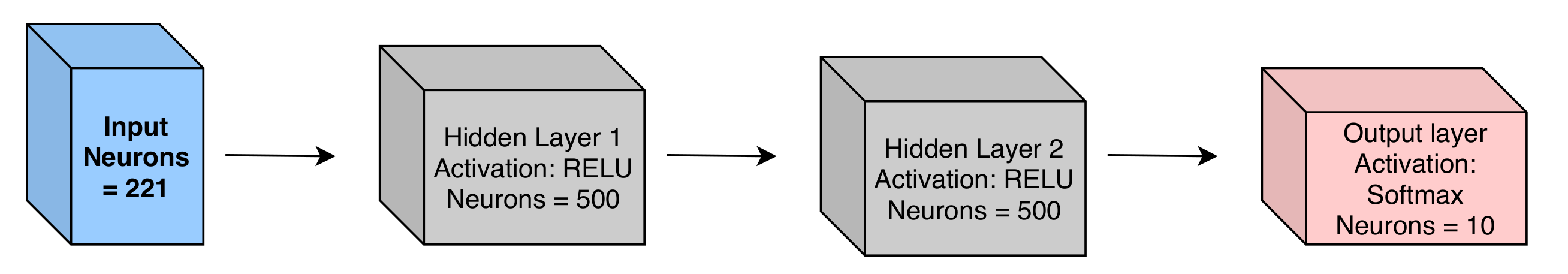}\\
	\caption{
    Architecture of the proposed \ac{nn}, with the number of input neurons corresponding to $\zeta = 10, \xi = 20$
	}
	\label{fig:architecture}
\end{figure}

The architecture of an \ac{nn} model includes the number of layers, number of neurons per layer, and how these neurons are connected. This architecture determines how complex it will be to calculate the optimal values for a specific task. A \ac{nn} with more layers and neurons typically requires a larger dataset for training.
Our \ac{nn} is composed of the input layer, two hidden layers, and one output layer, as depicted in \cref{fig:architecture}.

For \ac{nn} approaches, it is essential to define which features of the environment will be relevant to an effective solution. We use these features as the inputs of our model so that it may accurately react to the conditions of the environment. Our objective for the \ac{nn} model is to have it identify which of the \acp{bs} in $\Phi$ will provide the highest \ac{sinr} when connected via the directional antenna. This corresponds to a classification problem, wherein the \ac{nn} is trained to choose from one of several discrete options, given a provided input. Let $\Phi_\zeta \subset \Phi$ denote the $\zeta$ closest \acp{bs} to the \ac{uav}; the \ac{nn} will choose the serving \ac{bs} from within this set, based on which \ac{bs} it believes to have the highest directional antenna \ac{sinr}.

The \ac{nn} takes several measurements relating to the \acp{bs} in $\Phi_\zeta$ as inputs. First, it takes the time-averaged received signal powers from each of the relevant \acp{bs}, as measured by the omni-directional antenna $\bold{P}_\zeta = \{p_1,p_2,...,p_\zeta\}$, where $p_i = p \mu(\phi_i) (r_{i}^2+\Delta \gamma^2)^{-\alpha_{t_i}/2}$. The signal is time-averaged to remove the multipath fading effects.
As the UAV has access to the position information of the BSs, the \ac{nn} also takes the horizontal distances $\bold{R}_\zeta = \{r_1,r_2,...,r_\zeta\}$ to the \acp{bs} in $\Phi_\zeta$ as inputs. 

Using this same position information, along with knowledge of its directional antenna, the \ac{nn} only accounts for interfering BSs within its directional antenna beamwidth by taking in the horizontal distances only to these \acp{bs}. We denote a $\zeta \times \xi$ matrix as $\bold{F_\zeta}$, where each row corresponds to one of the candidate \acp{bs}, and each column corresponds to one of the $\xi$ closest \acp{bs} that would cause interference for the \ac{uav} if it attempted to communicate with one of the candidate \acp{bs}. In other words, $\bold{F_\zeta}(i,j)$ represents the distance from the UAV to the $j$-th closest \ac{bs} belonging to $\Phi_{\mathcal{W}_i}$, which is the set of \acp{bs} within the area $\mathcal{W}_i$ illuminated by the \ac{uav} directional antenna when aligning towards \ac{bs} $i$. In the event that $\Phi_{\mathcal{W}_i}$ contains fewer than $\xi$ \acp{bs}, the remaining entries in the $i$-th row of $\bold{F_\zeta}$ are set to null values.

Finally, the \ac{uav} takes it own height above ground $\gamma$ as an input into the \ac{nn}.


The \ac{nn} training itself requires the fine-tuning of parameters related to the learning rate and convergence of the classification, known as the hyperparameters.
Hyperparameters are parameters chosen before the training process that can improve the learning process.

We detail our choices for the hyperparameters below:

\begin{itemize}

\item {Learning Rate}: is the amount by which the weights in an \ac{nn} model are updated. We set it to $10^{-5}$; with this value, the model does not overfit to our training data.

\item {Epoch}: is an iteration of the training process where the model is filled 
with all the elements of the training dataset. 
If a model is trained with too many epochs, it can overfit to the training data, while if a model uses too few epochs, it might not learn the necessary features to perform the classification. 
After testing several values, we set the number of epochs to 200.


\item{Optimiser}: is the function that modifies the weights of each neuron with the purpose of minimising the loss function. The loss function indicates how close the output of the model is to the expected result. The main objective of the learning process is to optimise the loss function, making the predicted output closer to the expected one without over-fitting to the training data. We choose the optimiser AdaMax because it has the feature of accelerating the search for the minimum value of the loss function and reducing oscillations. In addition, it is less sensitive to the choice of the hyper-parameters when compared to the \ac{Adam} optimizer. 
\end{itemize}

 \subsection{Simulation \& Training}
In supervised learning, for a model to learn it must first be trained with a set of labelled data, and then tested with a second set to evaluate its accuracy. To avoid overfitting the model this second dataset cannot be used in the training process itself. To generate our datasets, we simulate the environment described in the System Model section, with a random \ac{ppp} distribution of \acp{bs} $\Phi$, random building heights, and the \ac{uav} at $x_0$ at a random height $\gamma$. The simulation is carried out in the R statistical language, using the R Keras library \cite{keras} for the \ac{nn} architecture. We record the values of $\bold{P}_\zeta$, $\bold{R}_\zeta$, and $\bold{F}_\zeta$, as observed by the \ac{uav}. We then have the \ac{uav} iteratively align its directional antenna with each of the candidate \acp{bs} in $\Phi_\zeta$ and we measure the time-averaged directional antenna \ac{sinr}. The index number of the \ac{bs} with the highest \ac{sinr} is stored as the label. This process is repeated a number of times, with random \ac{bs} positions, \ac{uav} heights and building deployments, to populate our datasets. Having generated the two datasets we train our \ac{nn} model to infer through the chosen features which \ac{bs} the \ac{uav} should associate with, for a given set of environmental parameters.

 

\section{Numerical Results}
\begin{figure}
\centering
	\includegraphics[width=.45\textwidth]{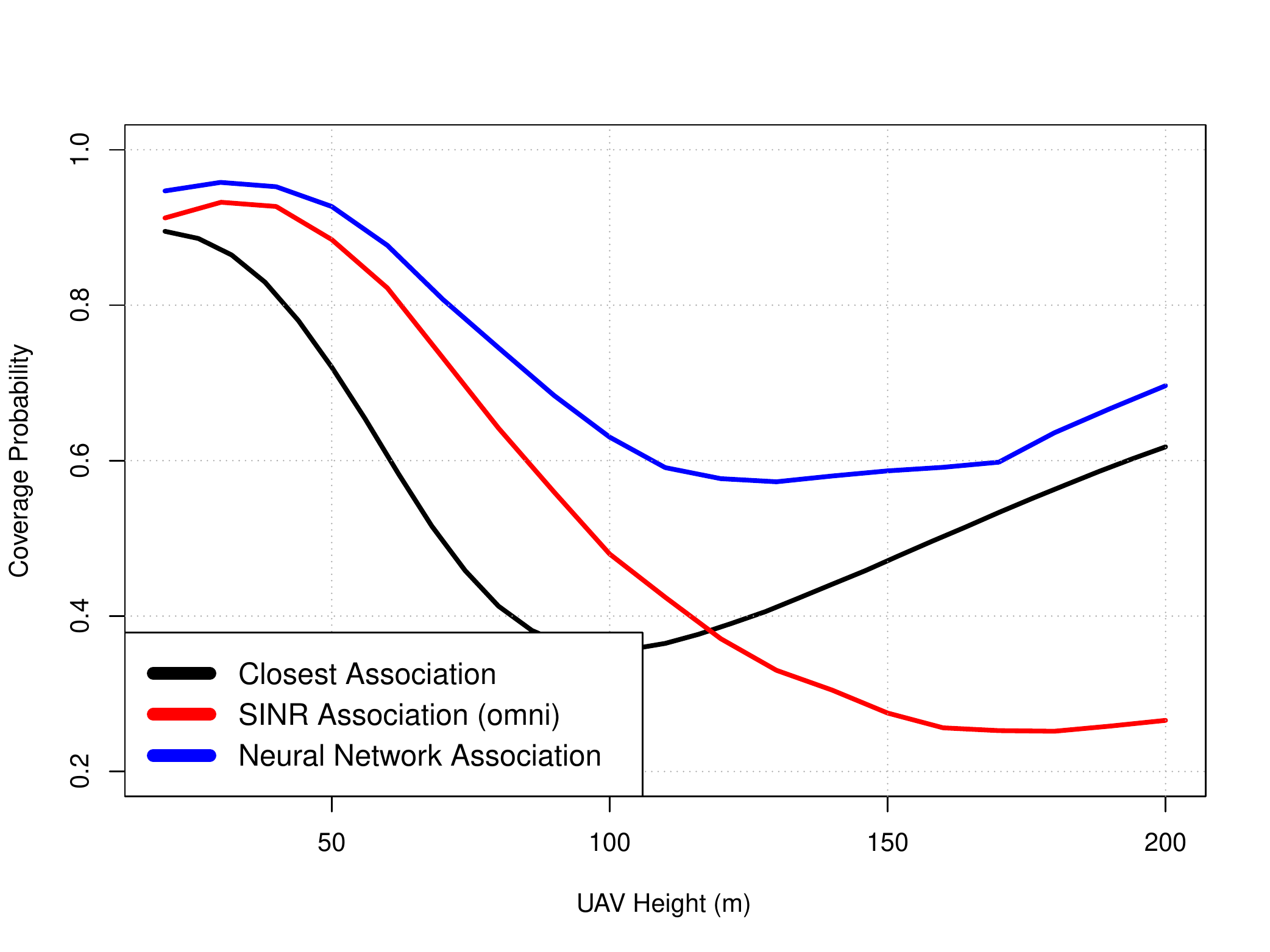}\\
	\vspace{-5mm}
	\caption{
	Coverage probability of the \ac{uav} as a function of the height $\gamma$, given beamwidth $\omega$ of 45 degrees and a \ac{bs} density $\lambda$ of $\unit[5]{/km^2}$. The blue line denotes the performance under our \ac{nn} association approach, the black line denotes the mathematically-derived performance for closest-\ac{bs} association derived in \cite{8422376} and \cite{Amer_2020}, and the red line denotes strongest \ac{sinr} association as measured from the omni-directional antenna.
	}
	\label{fig:height}
\end{figure}

\begin{figure}
\centering
	\includegraphics[width=.45\textwidth]{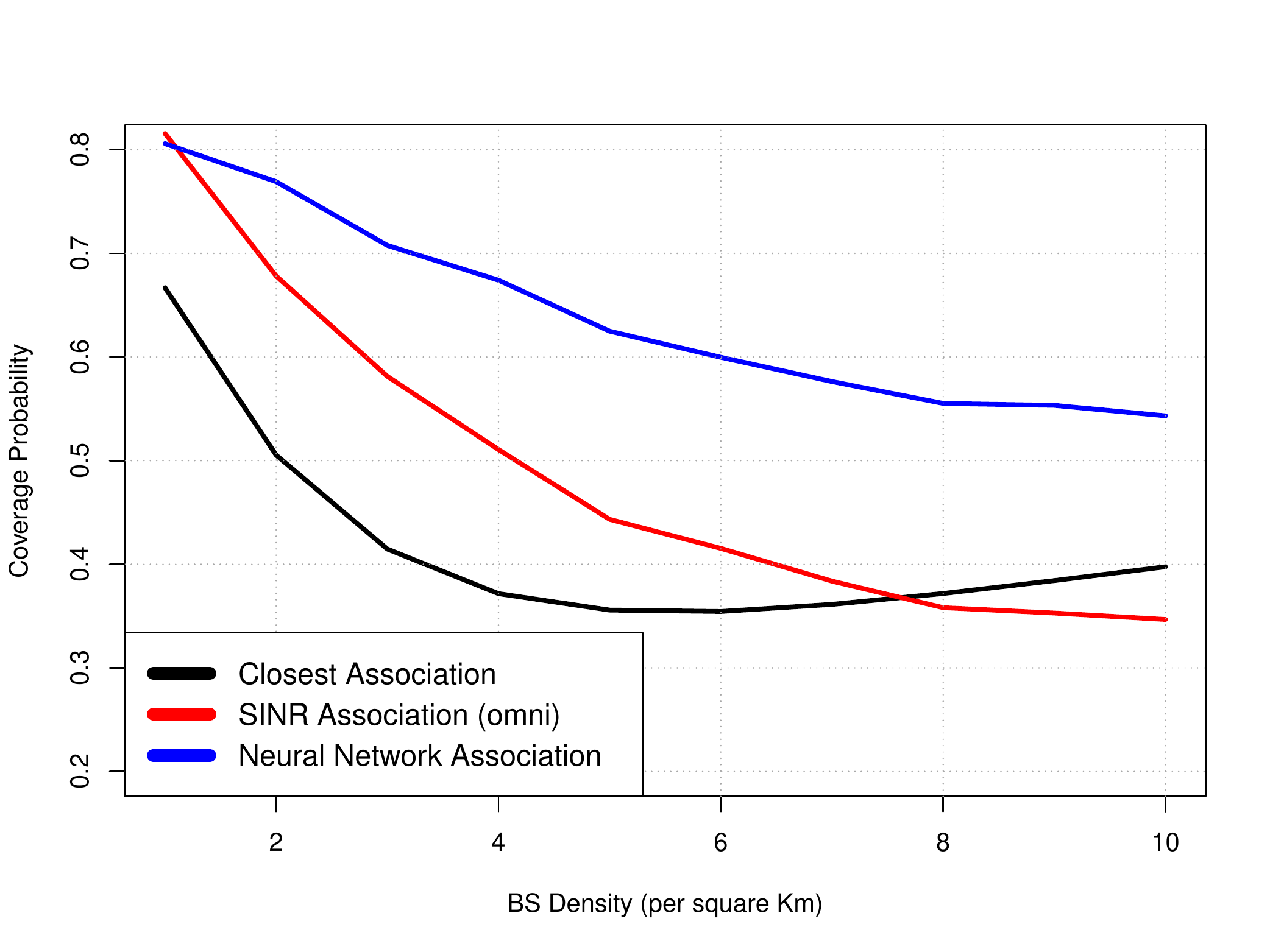}\\
	\vspace{-5mm}
	\caption{
  	Coverage probability of the \ac{uav} as a function of the \ac{bs} density $\lambda$, given beamwidth $\omega$ of 45 degrees and a \ac{uav} height $\gamma$ of $\unit[100]{m}$. The blue line denotes the performance under our \ac{nn} association approach, the black line denotes the mathematically-derived performance for closest-\ac{bs} association derived in \cite{8422376} and \cite{Amer_2020}, and the red line denotes strongest \ac{sinr} association as measured from the omni-directional antenna. 
	}
	\label{fig:density}
\end{figure}

\begin{figure}
\centering
	\includegraphics[width=.45\textwidth]{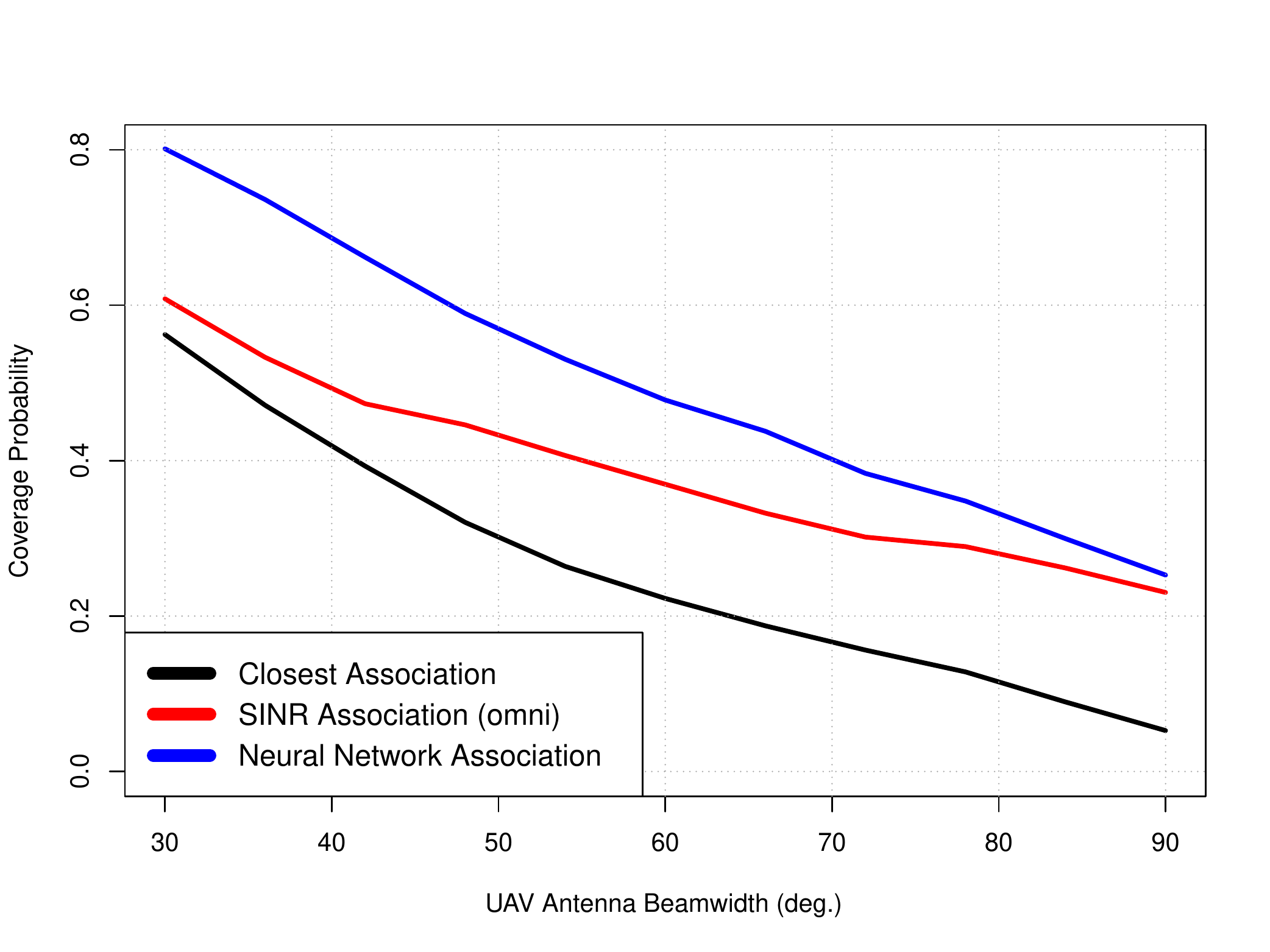}\\
	\vspace{-5mm}
	\caption{
  	Coverage probability of the \ac{uav} as a function of the \ac{uav} antenna beamwidth $\omega$, given \ac{bs} density $\lambda$ of $\unit[5]{/km^2}$ and a \ac{uav} height of $\unit[100]{m}$. The blue line denotes the performance under our \ac{nn} association approach, the black line denotes the mathematically-derived performance for closest-\ac{bs} association derived in \cite{8422376} and \cite{Amer_2020}, and the red line denotes strongest \ac{sinr} association as measured from the omni-directional antenna. 
	}
	\label{fig:Beamwidth}
\end{figure}

In this section we evaluate the performance of our \ac{nn}-based \ac{bs} association. We achieve this by simulating the urban environment over a number of \ac{mc} trials where the \ac{bs} distributions and the building heights are random, and recording the probability of the \ac{uav} having coverage after choosing a \ac{bs} to associate with (referred to as the coverage probability). For comparison, we additionally evaluate the performance of the \ac{uav} when it associates with a \ac{bs} following a simple heuristic scheme. As the \ac{uav} has an omni-directional antenna which monitors the \ac{bs} signal powers, the \ac{uav} can adopt an association scheme where it associates with the \ac{bs} which has the strongest \ac{sinr}, as directly measured by the omni-directional antenna. The \ac{uav} also knows the locations of the nearby \acp{bs}, therefore it can adopt an association scheme wherein it chooses to associate with the \ac{bs} that has the closest horizontal distance to it, irrespective of the received signal power at the omni-directional antenna. For both association schemes, the \ac{uav} makes a decision based on the information immediately available to it, and aligns the directional antenna towards its chosen associated \ac{bs}. 
Table 1 gives the values of the environmental parameters. In \cref{fig:height,fig:density,fig:Beamwidth}, the results for the closest-\ac{bs} association are obtained via the mathematical expressions derived by us in our prior works \cite{8422376} and \cite{Amer_2020}, while the rest are found via simulations.

\begin{table}[t!]
\vspace{-3mm}
\begin{center}
\caption{Numerical Result Parameters}
\begin{tabular}{ |c|c| } 
 \hline
 Parameter & Value \\ 
 \hline
 Carrier Frequency & \unit[2]{GHz} \\
  Building density & \unit[300]{$/\text{km}^2$}\\
 Building land coverage & 0.5\\
 Building height scale parameter & \unit[20]{m}\\
 \ac{los} pathloss exponent $\alpha_L$ & 2.1\\
 \ac{nlos} pathloss exponent $\alpha_N$ & 4\\
 \ac{los} multipath fading parameter $m_L$ & 1 \\
 \ac{nlos} multipath fading parameter $m_N$ & 1 \\
\ac{bs} transmit power $p$ & \unit[40]{W}\\
 Near-field pathloss $c$ & \unit[-38.4]{dB} \cite{Elshaer_2016} \\
 \ac{sinr} threshold $\theta$ & \unit[0]{dB} \\
 Noise power $\sigma^2$ & \unit[$8\cdot10^{-13}$]{W} \cite{Elshaer_2016} \\ 
 \ac{bs} height above ground $\gamma_{G}$ &  \unit[30]{m}\\
 Number of \ac{bs} antenna elements $N_t$ & 8 \\
 Candidate \ac{bs} number $\zeta$ & 10\\
 Interfering \ac{bs} number $\xi$ & 20\\
 \hline
\end{tabular}
 \label{tab:table}
\end{center}
\end{table}
 
\begin{figure}
\centering
	\includegraphics[width=.45\textwidth]{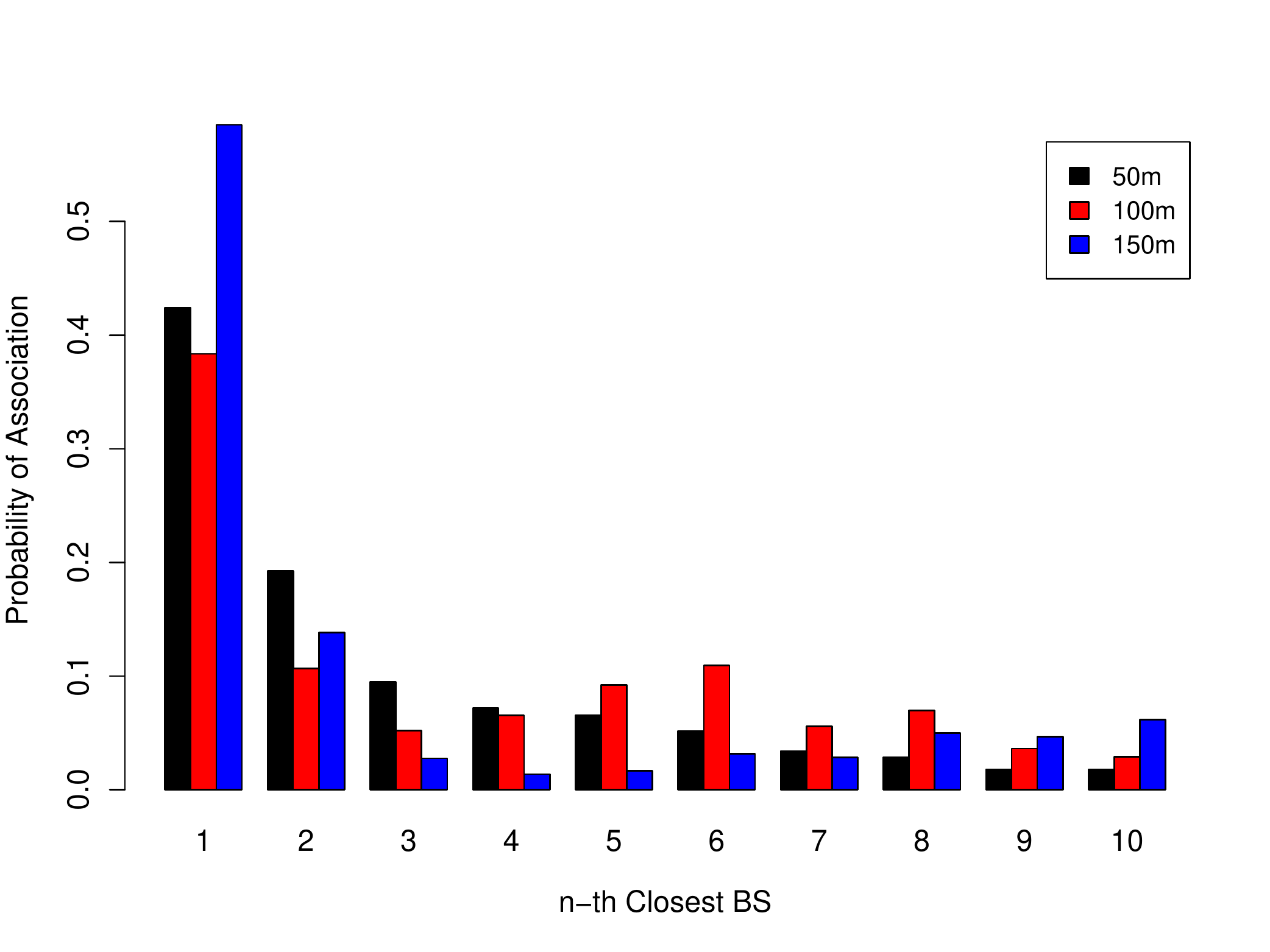}\\
	\vspace{-5mm}
	\caption{
	 Probability of the \ac{nn} choosing to associate to the $n$-th closest \ac{bs}, given a certain \ac{uav} height.
	}
	\label{fig:Association}
\end{figure}

The results in \cref{fig:height} show that the \ac{nn} association strategy gives a consistent improvement to the coverage probability when compared to the non-\ac{nn} closest-\ac{bs} or strongest \ac{sinr} association, at all values of the height of the \ac{uav}. At low heights the \ac{nn} strategy gives very similar performance to the strongest \ac{sinr} association, as interference effects are less of an issue, and therefore choosing the \ac{bs} with the strongest \ac{sinr} observed by the omni-directional antenna appears to be the optimum strategy. At large \ac{uav} heights, due to \ac{bs} antenna downtilt the stronger \ac{sinr} measured by the omni-directional antenna will come from more distant \acp{bs}, which will result in more interference for the directional antenna if the \ac{uav} associates with one of them (due to a shallower tilt angle and greater area $\mathcal{W}$). As a result, at large heights the \ac{uav} must prioritise connecting to a closer \ac{bs}, even if this will result in a lower \ac{bs} antenna gain, due to a bigger misalignment between the \ac{bs} antenna tilt, and the \ac{uav}-\ac{bs} vertical angle. The \ac{nn} recognises this, and so allows the \ac{uav} to massively improve its coverage probability over the strongest \ac{sinr} association case. The biggest \ac{nn} gains are achieved in the middle range of heights, where both the closest and strongest \ac{sinr} association strategies give poor performance. This is due to the fact that at these heights, on one hand the \ac{uav} has unobstructed \ac{los} channels to distant interfering \acp{bs}, while on the other hand the \ac{uav} is not so high up that it can mitigate interference through tilting its antenna down. The \ac{nn} is able to reduce the performance loss at these heights by choosing a \ac{bs} which offers a good tradeoff between a high sidelobe antenna gain, low signal pathloss, as well as low interference. 

The results in \cref{fig:density} show that the overall network performance deteriorates as the \ac{bs} density increases, due to increasing interference. At lower densities better performance is achieved by associating to the \ac{bs} with the strongest measured \ac{sinr}, while at larger densities connecting to the closest \ac{bs} gives better results. The \ac{nn} is able to outperform both association strategies, with a bigger performance improvement observed for the higher \ac{bs} densities where the \ac{uav} experiences more interference.

In \cref{fig:Beamwidth} we show the impact of the \ac{uav} directional antenna beamwidth $\omega$ on the coverage probability. Increasing the beamwidth causes performance to deteriorate for all association policies, due to the resulting increase in interference observed by the \ac{uav}. Note that as the beamwidth increases the peformance gain offered by our \ac{nn} solution decreases compared to the strongest association, as at larger beamwidths the intelligence of the \ac{nn} is not sufficient to mitigate the impact of interference.

The plot in \cref{fig:Association} shows the probability of the \ac{nn} choosing a certain \ac{bs} to associate with, for different \ac{uav} heights. We can see that the \ac{uav} will be served by the closest \ac{bs} approximately half of the time for the tested heights, due to the impact of antenna misalignment when the \ac{uav} and the \ac{bs} are a short horizontal distance apart. The \ac{nn} will instead sometimes prefer to connect to \acp{bs} further away, with the shape of the \ac{bs} sidelobes having a noticeable impact on the probability distribution of the chosen \ac{bs}. Consider, for example, how the \ac{nn} will very rarely choose the third-closest \ac{bs} when the \ac{uav} is at 100 meters; the \ac{nn} has learned during training that the fourth, fifth and sixth \acp{bs} are more likely to give a better \ac{sinr}, even if the distance-dependent pathloss and the interference is greater.

\section{Conclusion}
In this paper we have proposed an \ac{nn}-based association policy that allows a \ac{uav} to choose a suitable \ac{bs} to connect to, based on information about nearby \ac{bs} transmit powers, their distances to the \ac{uav}, as well as the locations of nearby \acp{bs} which may cause interference. We demonstrated that the resulting \ac{nn} was able to increase the probability of \ac{uav} coverage significantly, compared to typical non-\ac{nn} \ac{bs} selection schemes. 

In our future work we plan to extend the work in this paper by considering a scenario where the \ac{uav} moves through the urban environment, while making \ac{bs} association decisions. This will introduce handovers to the \ac{uav} association problem, which will complicate the decision process. We intend to apply machine learning to balance the \ac{uav} channel quality requirements with the additional mobility management requirements.

\section{Acknowledgement}
This material is based upon works supported by the Sci-
ence Foundation Ireland under Grants No. 17/NSFC/5224 and
13/RC/2077.

\ifCLASSOPTIONcaptionsoff
  \newpage
\fi



\bibliographystyle{./IEEEtran}
\bibliography{./IEEEabrv,./IEEEfull}
\end{document}